\documentclass[11pt]{article}

\usepackage{amsfonts, amsmath, graphics, graphicx, amssymb}
\usepackage{epsfig, psfig}

\newcommand{\pf}{\noindent {\bf Proof.}\hspace*{0.3cm}}
\newcommand{\epf}{\vspace*{-0.5cm}\hfill \framebox(6,6)[br] \\
  \vspace*{0.2cm}}

\newtheorem{Proposition}{Proposition}

\begin{document}

\date{}

\title{\textbf{On the Stability the Least Squares Monte Carlo}}
\author{
Oleksii Mostovyi\thanks{%email: omostovy@andrew.cmu.edu\\
Department of Mathematical Sciences, Carnegie Mellon University,
Pittsburgh, PA, 15213. } }

\maketitle
\begin{abstract}
Consider Least Squares Monte Carlo (LSM) algorithm, which is
proposed by Longstaff and Schwartz (2001) for pricing American
style securities. This algorithm is based on the projection of the
value of continuation onto a certain set of basis functions via the least
squares problem. We analyze the stability of the algorithm when
the number of exercise dates increases and prove that, if the
underlying process for the stock price is continuous, then the
regression problem is ill-conditioned for small values of
the time parameter.
\end{abstract}


\section{\textbf{Introduction}}

The problem of pricing American and Bermudan style options is of fundamental
importance in option pricing theory. In the continuous time settings McKean (1965) proposed an algorithm for
pricing an American put option with an infinite maturity via Ordinary
Differential Equations and
Partial Differential Equations. Further developments of this technique and applications to
other American style securities are considered in Peskir and
Shiryaev (2006). Another approach is pricing via Monte-Carlo
simulations that is described by Glasserman (2004). One of the
most difficult tasks in the theory of pricing of American and
Bermudan options is the determination of an optimal stopping rule
and the valuing of the option under such a rule. Longstaff
and Schwartz (2001) proposed an algorithm for pricing American and
Bermudan style options via Monte Carlo simulations, Least Squares
Monte Carlo or LSM. This technique is especially useful when we
deal with multi-factor processes. In this case the methods
based on binomial, trinomial trees, or partial differential equations become slow and thus
inefficient due to the high dimensionality of the problem.

As in the majority of the numerical algorithms the starting point of LSM for
American options is a substitution of the
continuous time interval with a discrete set of exercise dates.
Practically, by doing this we substitute the American option with
a Bermudan one. Then for each exercise time (except the first and
the last one) we project the value of continuation onto a set of
basis functions via linear regression. Clement, Lamberton
and Protter (2002) investigated the convergence of the algorithm
with the growth of the number of the basis functions and the Monte Carlo
simulations. Under fairly general conditions they proved the almost sure
convergence of the complete algorithm. Also, they obtained the rate of
convergence when the number of Monte Carlo simulations increases and showed
that the normalized error of the algorithm is asymptotically
Gaussian. However, they considered a fixed partition of the time
interval and thus, essentially, they discussed the properties of
the Bermudan, not American option.
Glasserman and Yu (2004)
investigated the behavior of LSM with the simultaneous grows of the
number of the basis functions and the number of the Monte-Carlo simulations
and estimated the rate of convergence in some more specific settings. Moreno and Navas (2001) considered the LSM
for different basis functions, namely, power series, Laguerre, Legendre,
Chebyshev polynomials, and deduced that the algorithm converges at
least for American Put options when the underlying problem has
a small number of factors. Stentoft (2004) obtained the rate of
convergence of the algorithm in the two period multidimensional
case.

In the present work we consider the stability of LSM algorithm, when the number of
exercise dates increases in such a way that there are exercise dates close to
an initial time, which we assume to be equal to zero without loss of generality. We prove that the algorithm is unstable
when the time parameter is close to zero, because the  underlying regression
problem is ill-conditioned.

The remainder of this work is organized as follows. In Section
\ref{description}, we describe the algorithm. In Section \ref{ill-conditioning},
we prove the main result, which is formulated in Proposition \ref{mainProp},
instability of the algorithm for the small values of the time parameter
due to the ill-conditioning of the corresponding matrix
in the regression problem. In addition  we present the results of the numerical
simulations that illustrate the assertions of Proposition \ref{mainProp}. In
Section \ref{conclusion}, we give the concluding remarks.

\section{\textbf{Description of the Algorithm}}\label{description}

 Assume that the stock price process $X=\{X_t\}_{t\in[0, T]}$ is given by the
 strong soluton to the following Stochastic Differential Equation:
\begin{equation}\label{sde}
\begin{array}{rcl}
dX_t & = & \mu_tdt + \sigma_tdW_t,\\
 X_0 & = & x_0.\\
\end{array}
\end{equation}
Here $x_0\in\mathbb{R}_{++}$ is a constant,
$W=\left\{W_t\right\}_{t\in[0, T]}$ is a Brownian Motion  (possibly
multidimensional)
 on a filtered probability space
$\left(\Omega, \mathcal{F}, \left\{\mathcal{F}_t\right\}_{t\in[0, T]},
  \mathbb{P}\right),$ where $\mathbb{P}$ is the
risk-neutral probability measure, $\mu$
and $\sigma$ are the progressively measurable functionals, such that the strong solution to equation
(\ref{sde}) exists and unique on the time interval $[0,~T]$, see Karatzas and
Shreve \cite{KaratzasShreve}, Chapter 5, for
the discussion of this topic.
The time horizon is $T,$ which we assume to be a finite constant. Usually equations of such a form are used to describe the evolution of the stock prices in
practice. Let the payoff of an American option at the
time of exercise $\tau$ is given by $\varphi\left(X_\tau\right),$ where
$\varphi$ is the corresponding payoff function.
Then the value of the option is determined by the formula:
\begin{equation}\nonumber
\nu(X_0)\triangleq \sup\limits_{\tau\in\mathcal{A}}\mathbb{E}\left[e^{-\int_{0}^{\tau}r_udu}\varphi(X_{\tau})\right].
\end{equation}
Here $r=\left\{ r_t\right\}_{t\in[0, T]}$ is an interest rate process, which we assume to
be deterministic for simplicity; $\mathcal{A}$ is the set
of the stopping times with
respect to the filtration $ \left\{\mathcal{F}_t\right\}_{t\in[0, T]}.$

To
approximate $\nu(X_0)$ numerically let us conduct $N$ Monte-Carlo
simulations of the process $X$. First, we need to divide the
time interval $[0,~ T]$ into $M$ subintervals $[t_m,
~t_{m+1}]$ of the length $\triangle t \triangleq \frac{T}{M}$, where
$t_m\triangleq\frac{T}{M}m$,\@ $m=1, ~..., ~M$. Thus at every moment $t_m$
we obtain $N$ realizations of the process $X^n_{t_m}$, $n=1,
..., N$. Second, for each simulation we compute the value of the option
at time $t_M=T$ (under the assumption that the option was not exercised before $T$):
\begin{equation}\nonumber
C^n_{t_M}\triangleq \varphi(X^n_{t_M}),~~~n=1, ~..., ~N.
\end{equation}
Discounting these values we get a cash flow vector
\begin{equation}\nonumber
b_{M-1}\triangleq d\left(C^1_{t_M},~ ..., ~ C^N_{t_M}\right)^T
\end{equation}
where $d\triangleq exp(-\int_{t_{M-1}}^{t_M}r_udu)$ is the discount factor.

To obtain the value of the option at $t_{M-1},$ $C_{t_{M-1}}$ (under the assumption the option was not
exercised before $t_{M-1}$), we chose a hypothesis of
linear regression and project the cash flow vector $b_{M-1},$ for example, on a constant,
$X_{t_{M-1}},$ and $X^2_{t_{M-1}}$. According to \cite{LSM} this is one of the simplest yet
successful regression models. According to \cite{Moreno} a good alternative
choice of basis functions can be Hermite, Laguerre,
Legendre, or Chebyshev polynomials. If we use $1, X_{t_{M-1}},$ and $X_{t_{M-1}}^2$ as the basis, the
estimate of the conditional expectation becomes
\begin{equation}\label{condExpReg}
\mathbb{E}[C_{t_{M-1}}|\mathcal{F}_{t_{M-1}}]= \alpha + \beta X_{t_{M-1}}  + \gamma
X^2_{t_{M-1}},
\end{equation}
where $\alpha$,\@ $\beta$,\@ $\gamma$ are some constants.
Then along every path we compare values of immediate exercise,
$\varphi(X^n_{t_{M-1}})$, with values of continuation that are
obtained by substitution of $X^n_{t_{M-1}}$ into equation
(\ref{condExpReg}). The bigger of two gives
$C^n_{t_{M-1}}$, $n = 1, ~..., ~N$. If value of immediate exercise
is bigger we set $C^n_{t_M}=0$.

Similarly we obtain $\mathbb{E}\left[C_{t_m}|\mathcal{F}_{t_m}\right]$ for each
$m\in\left\{M-1, \dots, 1 \right\}$ via solving
linear regression problems $A(t_m)x(t_m)=b(t_m),$ where components of the
matrix $A(t_m)$ depend on the regression hypothesis and the outcomes of
the simulations, $x(t_m)$ is an unknown
vector of coefficients, and the vector $b(t_m)$ is given by equation
\begin{equation}\label{cashFlowVector2}
b_{t_m}=d\left(C^1_{t_{m+1}},~ ..., ~ C^N_{t_{m+1}}\right),
\end{equation}
where $d=exp\left(-\int_{t_m}^{t_{m+1}}r_udu\right)$ is the discount factor.

 Finally we discount the cash flow up
to the moment of time $t_0=0$ and compare it with the value of
immediate exercise at time $t_0$,\@ $\varphi(X_{t_0})$. The bigger
is the value of the option.

\section{\textbf{Ill-Conditioning for small $t$}}\label{ill-conditioning}
Let $\mathcal P = \left\{ t_m= \frac{T}{M}m: \hspace{2mm} m=1, \dots, M-1
\right\}$ be a partition of the interval $[T/M, T(M-1)/M].$ In order to
compute the estimates of the value of the option at each $t\in \mathcal P$
(under the assumption that it was not exercised before) we solve the linear regression problem
\begin{equation}\label{regression}
 A(t)x(t)=b(t),
\end{equation}
where $x(t)$ is an unknown vector of coefficients, vector $b(t)$ is
determined by equation (\ref{cashFlowVector2}), and the matrix
$A(t)$ depends on the regression hypothesis and the outcome of the Monte Carlo
simulations. Assume that we have
chosen $K$ continuous functions $f_1, \cdots, f_K$ as the
hypothesis. Examples of such functions are power series, Laguerre,
Legendre, Hermite polynomials, etc. In this case $A(t)$ has
the following form
\begin{equation}\label{matrixA}
 A(t)\triangleq \left( \begin{array}{ccc}
   f_1\left(X^1_{t}\right) & \dots & f_K\left(X^1_{t}\right) \\
   & \dots &   \\
   f_1\left(X^N_{t}\right) & \dots & f_K\left(X^N_{t}\right) \\
 \end{array} \right),
\end{equation}
where $N$ is the number of the simulations.
We show that, if the underlying process $X$ is almost surely continuous,
then for small $t$ the problem (\ref{regression}) is
ill-conditioned. For a matrix $A$ in the $l_2$ norm the \textbf{condition number} is
defined as
\begin{equation}\label{defKappa}
\kappa(A) \triangleq || A ||\cdot|| A^{-1} || = \frac{\sigma_{max}(A)}{\sigma_{min}(A)},
\end{equation}
where $\sigma_{max}(A)$ and $\sigma_{min}(A)$ are maximal and minimal
singular values of the matrix $A$ respectively.
A problem with a low condition number is called \textbf{well-conditioned},
while a problem with a high condition number is called
\textbf{ill-conditioned}.

Usually, problem (\ref{regression})  is solved via one of the following methods:
Householder triangularization, Gram-Schmidt orthogonalization,
singular value decomposition, or normal equations. Let $\kappa$ be
the condition number of matrix $A$.
If $\left(A^TA\right)^{-1}$ exists then the exact solution to the Least-Squares Problem is given by the vector $x=\left(A^TA\right)^{-1}A^Tb,$
i.e. it is a product of the left-inverse of the matrix $A$ and the vector
$b.$ One can see that the solution  to the problem
(\ref{regression}), obtained via normal equations, is governed
by $\kappa ^2$, whereas the solution obtained via SVD, Householder or Gram-Schmidt is governed by $\kappa$. Consequently, normal
equations are the least stable with respect to the grows of the condition number.
Nevertheless, the analytical solution to the Least-Squares Problem is defined
in terms of the normal equations.

Let $\kappa(t)$ denote the condition number of the matrix $A(t)$ for  $t\in[0, T].$ We show below that $\lim\limits_{ t\downarrow 0 } \kappa(t)
= \infty$ $\mathbb{P}$-almost surely. Therefore, no matter what algorithm one uses
(Householder, Gram-Schmidt, SVD or normal equations), for small values of time
the underlying
regression problem is ill-conditioned, and thus the algorithm is unstable.

We prove ill-conditioning for an arbitrary number of basis functions
in the following proposition. In addition, we illustrate the
phenomenon with the results of the numerical simulations in the case
when regression is done on three basis functions $1$, $x$, and
$x^2.$

\begin{figure}[h]
\begin{center}

 \includegraphics[width=0.9\textwidth]{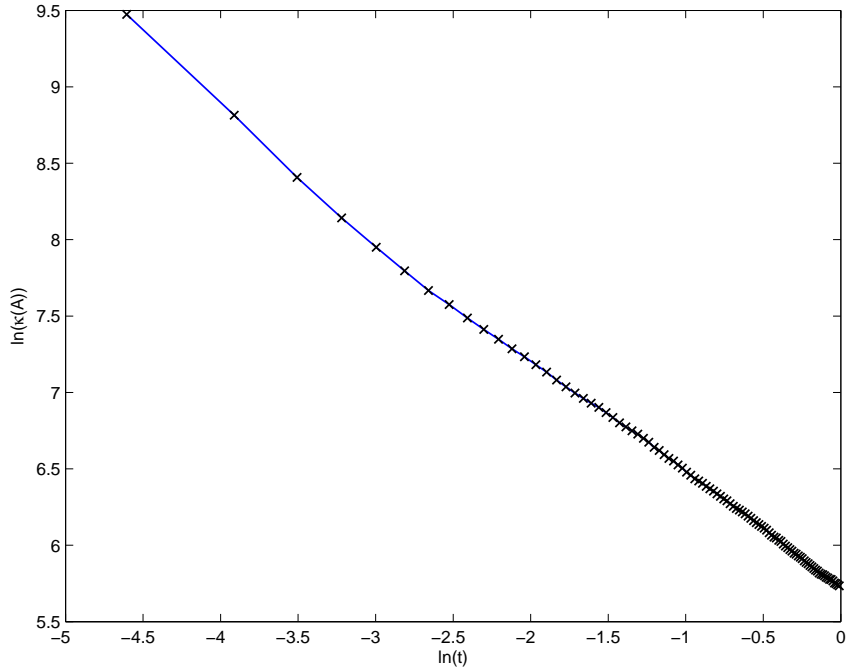}
 \caption{$ln\left(\kappa(A)\right)$ as a function of $ln(t),$ $t\in(0.009,
   1],$ $X$ - lognormal process, $\mu\equiv 0,$ $\sigma\equiv 0.15,$ \@$30000$ paths,
   $100$ time steps, Milstein discretization scheme.}
\end{center}
\end{figure}

\begin{figure}[h]
 \centerline{
  \includegraphics[width=0.9\textwidth]{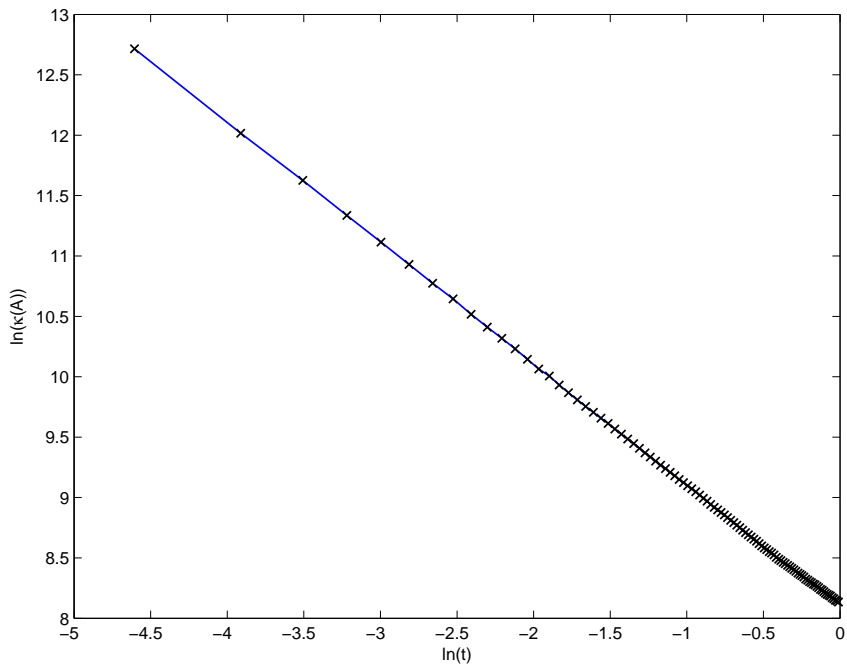}
    }
  \caption{$ln\left(\kappa(A)\right)$ as a function of $ln(t),$ $t\in(0.009,
    1],$ $X$ - normal process, $\mu\equiv 0,$ $\sigma\equiv 0.03,$ $30000$ paths, $100$
    time steps, Euler discretization scheme.}
  \label{figNormal}
  %\end{center}
\end{figure}

\begin{Proposition}\label{mainProp}
Assume that the process $X$ is given by equation (\ref{sde}), $f_k$,
$k=1, ~..., ~K$, are continuous functions, such that $\sum_{k=1}^K
f^2_k\left( X_0 \right) > 0.$
 Let for each $t\in[0, T]$ the matrix $A(t)$ is defined by
equation (\ref{matrixA}) and $\kappa(t)$ is the condition number of
$A(t).$ Then
\begin{equation}\label{conditioning}
\mathbb P\left[ \lim\limits_{t\rightarrow 0^+} \kappa (t) = \infty \right]
= 1.
\end{equation}
\end{Proposition}
\pf Consider equation (\ref{matrixA}). It follows from equation (\ref{sde})
that all rows of $A(0)$ are
identical. Consequently the rank of $A(0)$ equals to $1$. Let us look at
the following matrix:
\begin{equation}\label{AtrAat0}
\frac{(A^TA)(0)}{N}=\left(
\begin{array}{cccc}
f^2_1(X_0)       & f_1(X_0)f_2(X_0) & \dots & f_1(X_0)f_K(X_0) \\
f_2(X_0)f_1(X_0) & f^2_2(X_0)       & \dots & f_2(X_0)f_K(X_0) \\
                &                  & \dots &                  \\
f_K(X_0)f_1(X_0) & f_K(X_0)f_2(X_0) & \dots & f^2_K(X_0)       \\
\end{array}
\right).
\end{equation}
The matrix $\left(A^TA\right)(0)/N$ has two eigenvalues:
$\sum_{k=1}^K f^2_k(X_0)$ is the first one (with multiplicity $1$),
 $0$ is the second one (with multiplicity $K-1$). Thus $A(0)$
 has singular values $\sqrt{N\sum_{k=1}^K f^2_k(X_0)}$ and $0,$ consequently
$\kappa(0)=\infty$.

Note that the matrix $\left(A^TA\right)(t)$ is real and symmetric,
thus the eigenvalues of $\left(A^TA\right)(t)$ are real for all
$t'$s. Since eigenvalues are continuous functions of the components
of the matrix and the components in turn are a.s. continuous
processes, we deduce that for small $t'$s the first eigenvalue is in
the neighborhood of $\sum_{k=1}^K f^2_k(X_0)N,$ which is greater
then zero by the assumption of proposition, whereas all other
eigenvalues are is in the neighborhood of $0$. The conclusion,
$\mathbb P\left[ \lim\limits_{t\rightarrow 0^+} \kappa (t) = \infty
\right] = 1$ follows from continuity of the underlying process $X$
and the basis functions $ f_k $'s, $1\leq k\leq K.$

\epf

\vspace{1mm}

Intuitively Proposition \ref{mainProp} shows that for small values of $t$ the
condition number is large.

\section{\textbf{Concluding Remarks}}\label{conclusion}

We proved that for a continuous underlying stock price process,  LSM
algorithm for pricing American options is unstable when time parameter is
small. An interesting question is to obtain an exact bound on
applicability of this algorithm. A possible criterion of
applicability is the condition number of matrix $A(t)$, $\kappa
(t)$. For example, if $\kappa (t)$ exceeds a certain value, one can
treat (\ref{regression}) as a rank deficient least squares problem (see
\cite{GolubVanLoan} for details), or switch to another method:
backward induction or the method introduced by McKean \cite{McKean} of option
pricing via Ordinary Differential Equations or Partial Differential Equations
considered on a smaller domain. For certain problems it is
possible to obtain desired accuracy using relatively small number
of time intervals, then one does not have to solve the regression
problem for small $t'$s, and consequently the algorithm can be
stable. Also, if the underlying process $X$ is discontinuous
with high probability the algorithm can be stable even for small
values of the time parameter.

\end{document}